\documentclass{desyproc}

\begin{document}
\title{Beauty baryon production in $pp$ collisions at LHC and $b$
quark distribution in proton}

\author{{\slshape G.I.Lykasov$^1$, V.V.Lyubushkin$^1$,
T.V.Lyubushkina$^1$ and V.A.Bednyakov$^1$}\\[1ex]
$^1$JINR, Dubna, 141980, Moscow region, Russia
}

\contribID{smith\_joe}

\desyproc{DESY-PROC-2008-xx}
\acronym{EDS09}

\maketitle

\begin{abstract}
 The production of charmed and beauty hadrons
 in proton-proton and proton-antiproton collisions at high
energies are analyzed within the modified quark-gluon string model
(QGSM) including the internal motion of quarks in colliding 
hadrons. 
 We present some predictions for the future experiments on the
beauty baryon production in $pp$ collisions at LHC energies. This analysis
allows us to find interesting information on the Regge trajectories of 
the heavy ($b{\bar b}$) mesons and the sea beauty quark distributions in 
the proton.

\end{abstract}

\section{Introduction}
Various approaches of perturbative QCD including the
next-to-leading order calculations (NLO QCD) have been applied to
construct distributions of quarks in a proton. The theoretical
analysis of the lepton deep inelastic scattering (DIS) off protons
and nuclei provides rather realistic information on the
distribution of light quarks like $u,d,s$ in a proton. However, to
find a reliable distribution of heavy quarks like $c({\bar c})$
and especially $b({\bar b})$ in a proton describing the
experimental data on the DIS is a non-trivial task. 
It is mainly
due to small values of $D$ and $B$ meson yields in the DIS at
existing energies. Even at the Tevatron energies the $B$- meson
yield is not so large. At LHC energies the multiplicity of these
mesons produced in $pp$ collisions will be significantly
larger. Therefore one can try to extract a new information on the
distribution of these heavy quarks in a proton. In this paper we
suggest to study the distribution of heavy quarks like $c({\bar
c})$ and $b({\bar b})$ in a proton from the analysis of the future
LHC experimental data.

The multiple hadron production in hadron-nucleon collisions at
high energies and large transfers is usually analyzed
within the hard parton scattering model (HPSM) suggested in 
\cite{Efremov,FF}. This model was applied to
the charmed meson production both in proton-proton and meson-proton 
interactions at high energies, see for example \cite{Bednyakov:1995}. 
The HPSM is significantly improved by applying the QCD parton approach 
\cite{Nasson,Kniehl2}, see details in \cite{LKSB:09} and references
therein.
Unfortunately the QCD including the next-to-leading order (NLO) 
has some uncertainties related to the renormalization parameters especially at 
small transverse momenta $p_t$ \cite{LKSB:09}. 

In \cite{LKSB:09,LLB:09} we studied the charmed and beauty meson production
in $pp$ and $p{\bar p}$ collisions at high energies within the QGSM 
\cite{kaid1} or the dual parton model (DPM) \cite{Capella:1994} based on the $1/N$ 
expansion in QCD \cite{tHooft:1974,Veneziano:1974}.   
It was shown that this approach can be applied rather successfully at not very large values
of $p_t$.
In this paper we investigate the open charm and beauty baryon production
in $pp$ collisions at LHC energies and very small $p_t$ within the QGSM to find
new information on the Regge trajectories of the
heavy ($c{\bar c}$) and ($b{\bar b}$) mesons and the sea beauty quark distributions in the 
proton.

\section{General formalism for hadron production in $pp$ collision within QGSM}
Let us present briefly the scheme of the analysis of the hadron production in the $pp$ 
collisions within the QGSM including the transverse motion of quarks and diquarks in
colliding protons \cite{LAS}. As is known, the cylinder type
graphs for the $pp$ collision presented in Fig.1 make
the main contribution to this process \cite{kaid1}. 
The left diagram of Fig.1, the so-called
one-cylinder graph, corresponds to the case where two colorless
strings are formed between the quark/diquark ($q/qq$) and the
diquark/quark ($qq/q$) in colliding protons; then, after their
breakup, $q{\bar q}$ pairs are created and fragmentated to a hadron,
for example, $D$ meson. The right diagram of Fig.1, the
so-called multicylinder graph, corresponds to creation of the same
two colorless strings and many strings between sea
quarks/antiquarks $q/{\bar q}$ and sea antiquarks/quarks ${\bar
q}/q$ in the colliding protons.
\begin{figure}[htb]
\vspace{9pt}
\includegraphics[scale=1.0]{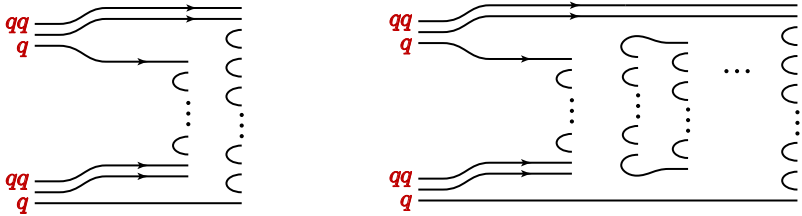}
\caption{The one-cylinder graph (left diagram) and the multicylinder 
graph (right diagram) for the inclusive $p p\rightarrow h X$ process.}
\label{Fig.1}
\end{figure}
The general form for the invariant inclusive hadron spectrum
within the QGSM is \cite{kaid2,LAS}
\begin{eqnarray}
E\frac{d\sigma}{d^3{\bf p}}\equiv
\frac{2E^*}{\pi\sqrt{s}}\frac{d\sigma}{d x d p_t^2}=
\sum_{n=1}^\infty \sigma_n(s)\phi_n(x,p_t)~, 
\label{def:invsp}
\end{eqnarray}
where $E,{\bf p}$ are the energy and the three-momentum of the
produced hadron $h$ in the laboratory system (l.s.) of colliding protons; 
$E^*,s$ are the energy of $h$ and the square of the initial energy in the
c.m.s of $pp$; $x,p_t$ are the Feynman variable and the transverse
momentum of $h$; $\sigma_n$ is the cross section for production of
the $n$-Pomeron chain (or $2n$ quark-antiquark strings) decaying
into hadrons, calculated within the ``eikonal approximation''
\cite{Ter-Mart}. Actually, the function $\phi_n(x,p_t)$ is the convolution
of the quark (diquark) distributions in the proton and their fragmentation functions
(FF) , see details in \cite{kaid1,Capella:1994,LKSB:09,LAS}.
To calculate the interaction function $\phi_n(x,p_t)$ we have to know all the quark (diquak)
distribution functions in the nth Pomeron chain and the FF. 
They are constructed within the QGSM using the knowledge of the secondary Regge trajectories,
see details in \cite{kaid1,kaid2}. 
\section{Heavy baryon production within QGSM}
\subsection{Sea charm and beauty quark distribution in proton}

Now let us analyze the charmed and beauty baryon production in the $pp$
collision at LHC energies and very small $p_t$ within the soft QCD, e.g.,
the QGSM. This study can be interesting for it may allow 
predictions for future LHC experiments like TOTEM and ATLAS and 
an opportunity to find new information on the distribution
of sea charmed ($c$) and beauty ($b$) quarks at very low $Q^2$. 
According to the QGSM, the distribution of $c({\bar c})$ quarks in the $n$th Pomeron 
chain (Fig.1(right)) is, see for example \cite{LAS} and references therein,
\begin{eqnarray}
f_{c({\bar c})}^{(n)}(x) = C_{c({\bar c})}^{(n)}\delta_{c({\bar c})}
x^{a_{cn}} 
(1-x)^{g_{cn}}
\quad
\label{def:fc}
\end{eqnarray}
where $a_{cn}=-\alpha_\psi(0)$, 
$g_{cn}=\alpha_\rho(0)-2\alpha_B(0)+(\alpha_\rho(0)-\alpha_\psi(0))+n-1$;
$\delta_{c({\bar c})}$ is the weight of charmed pairs in the quark sea, 
$C_{c({\bar c})}^{(n)}$
is the normalization coefficient \cite{kaid2},
 $\alpha_\psi(0)$ is the intercept of the $\psi$- Regge trajectory.
Its value can be $-2.18$ assuming that this trajectory $\alpha_\psi(t)$ 
is linear and the intercept and the slope $\alpha_\psi^\prime(0)$ can be
determined by drawing the trajectory through the $J/\Psi$-meson mass 
$m_{J/\Psi}\simeq 3.1$ GeV and the $\chi$-meson mass $m_\chi=3.554$ GeV 
\cite{Boresk-Kaid:1983}. Assuming that the $\psi$- Regge trajectory is 
nonlinear one can get  $\alpha_\psi(0)\simeq 0$, which follows from perturbative 
QCD, as it was shown in \cite{Kaid-Pisk:1986}. 
The  distribution of $b({\bar b})$ quarks in the $n$th Pomeron 
chain (Fig.1(right)) has the similar form  
\begin{eqnarray}
f_{b({\bar b})}^{(n)}(x) = C_{b({\bar b})}^{(n)}\delta_{b({\bar b})}
x^{a_{bn}}
(1-x)^{g_{bn}}
\quad
\label{def:fb}
\end{eqnarray}
where $a_{bn}=-\alpha_\Upsilon(0)$, $g_{bn}=\alpha_\rho(0)-2\alpha_B(0)+(\alpha_\rho(0)-\alpha_\Upsilon(0))+n-1$; 
$\alpha_\rho(0)=1/2$ is the well known intercept of the $\rho$-trajectory; $\alpha_B(0)\simeq -0.5$
is the intercept of the baryon trajectory, $\alpha_\Upsilon(0))$ is the intercept  of the 
$\Upsilon$- Regge trajectory, its value also has an uncertainty. Assuming its linearity 
one can get $\alpha_\Upsilon(0))=-8, -16$, while for nonlinear ($b{\bar b}$) Regge trajectory
$\alpha_\Upsilon(0)\simeq 0$, see details in \cite{Piskunova}.
Inserting these values to the form
for $f_{c({\bar c})}^{(n)}(x)$ and $f_{b({\bar b})}^{(n)}(x)$ we get the large sensitivity
for the $c$ and $b$ sea quark distributions in the $n$th Pomeron chain.
Note that the FFs also depend on the parameters of these Regge trajectories. Therefore,
the knowledge of the intercepts and slopes of the heavy-meson Regge trajectories is
very important for the theoretical analysis of open charm and beauty production in
hadron processes.

Note that all the quark distributions obtained within the QGSM are different from the parton
distributions obtained within the perturbative QCD which are usually compared with the experimental
data on the deep inelastic lepton scattering (DIS) off protons. To match these two kinds of quark 
distributions one can apply the procedure suggested in \cite{Cap-Kaid}.
The quantities $g_{cn}$ or $g_{bn}$ entering into Eq.(\ref {def:fc}) and Eq.(\ref {def:fb}) are 
replaced by the following new quantities depending on $Q^2$
\begin{equation}
{\tilde g}_{cn}=g_{cn}(1+\frac{Q^2}{Q^2+c})~;~ {\tilde g}_{bn}=g_{bn}\left(1+\frac{Q^2}{Q^2+d}\right)
\end{equation}
The parameters $c$ and $d$ are chosen such that the structure function constructed from the
valence and sea quark (antiquark) distributions in the proton should be the same as 
the one at the initial conditions at $Q^2=Q_0^2$ for the perturbative QCD evolution.A similar
procedure can be used to get the $Q^2$ dependence for the powers $a_{cn}$ and $a_{bn}$ entering
into Eqs.(\ref{def:fc},\ref{def:fb}) \cite{Cap-Kaid}. Then using the DGLAP evolution equation 
\cite{DGLAP} we obtain the structure functions at large $Q^2$. 
 \subsection{Charmed and beauty baryon production in $pp$ collision} 

The information on the charmonium ($c{\bar c}$) and botomonium ($b{\bar b}$) Regge trajectories
can be found from the experimental data on the charmed and beauty baryon production in $pp$ 
collisions at high energies. For example, Fig.2 illustrates the sensitivity of the inclusive
spectrum $d\sigma/dx$ of the produced charmed  baryons $\Lambda_c$ to different values for 
$\alpha_\psi(0)$. 
The solid line corresponds to $\alpha_\psi(0)=0$, whereas the dashed curve corresponds to 
$\alpha_\psi(0)=-2.18$.
Unfortunately the experimental data presented in Fig.2 have big uncertainties; 
therefore, one cant extract the information on the $\alpha_\psi(0)$ values from the 
existing experimental data.
\begin{figure}[ht]
   \begin{center}
 {\epsfig{file=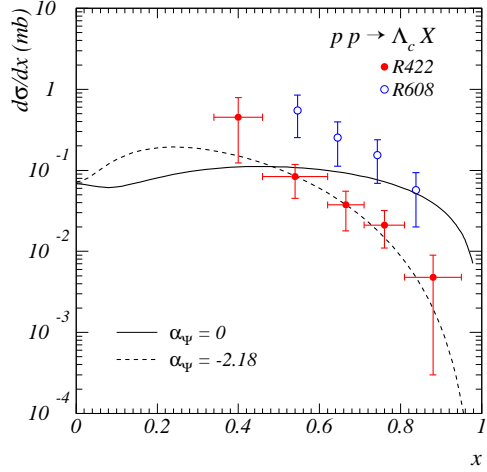,width=0.45\linewidth  }}
 \caption[Fig.2]{The differential cross section $d\sigma/dx$ for the
 inclusive process $pp\rightarrow\Lambda_c X$ at $\sqrt{s}=62~\mathrm{GeV}$.
 } 
   \end{center}
 \end{figure}
A high sensitivity of the inclusive spectrum $d\sigma/dx$ of the produced beauty baryons 
$\Lambda_b$ to different values for $\alpha_\Upsilon(0)$ is presented in Fig.3 (left).
\begin{figure}[htb]
\begin{center}
\begin{tabular}{cc}
\mbox{\epsfig{file=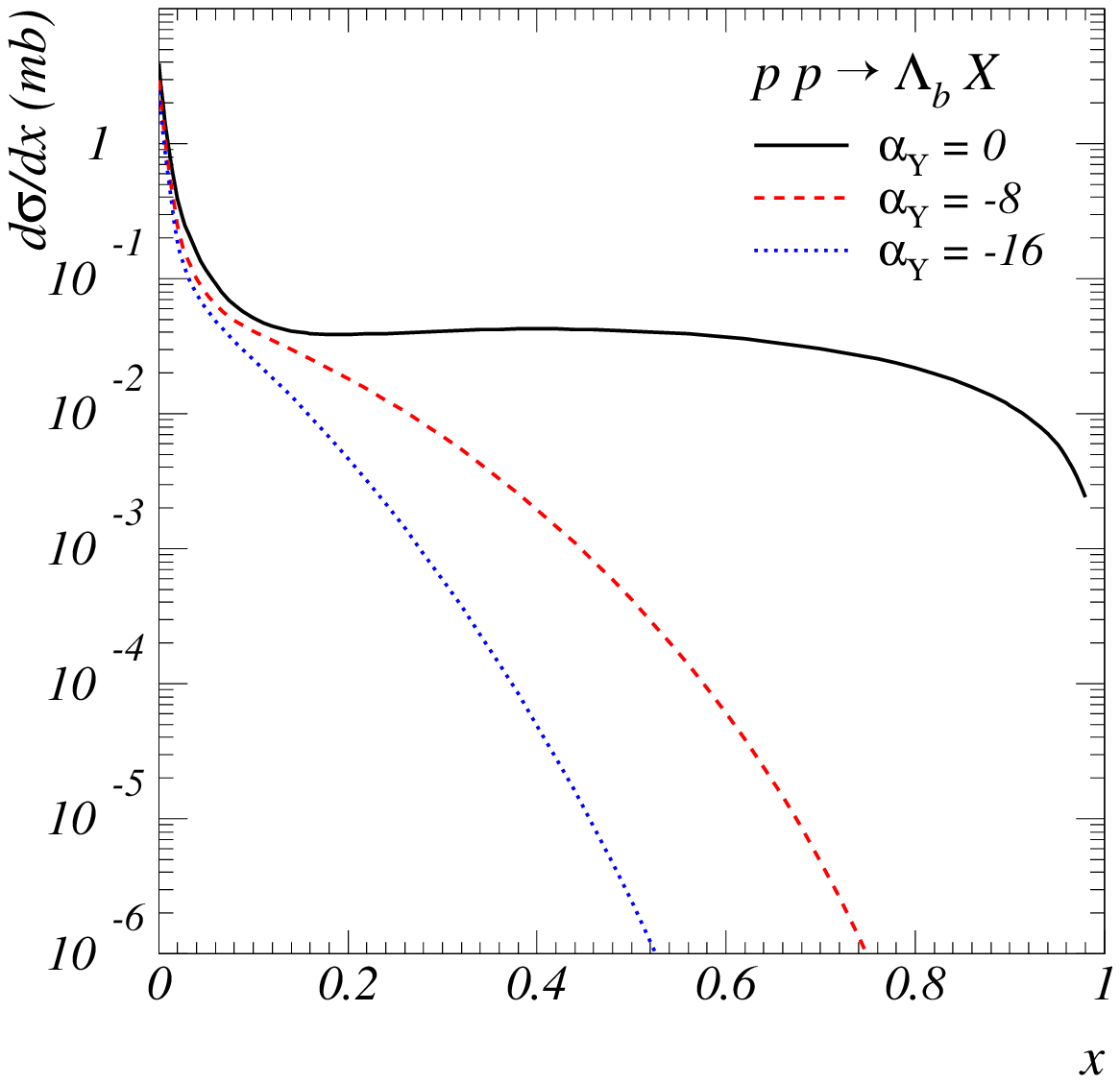,width=0.45\linewidth}} &
\mbox{\epsfig{file=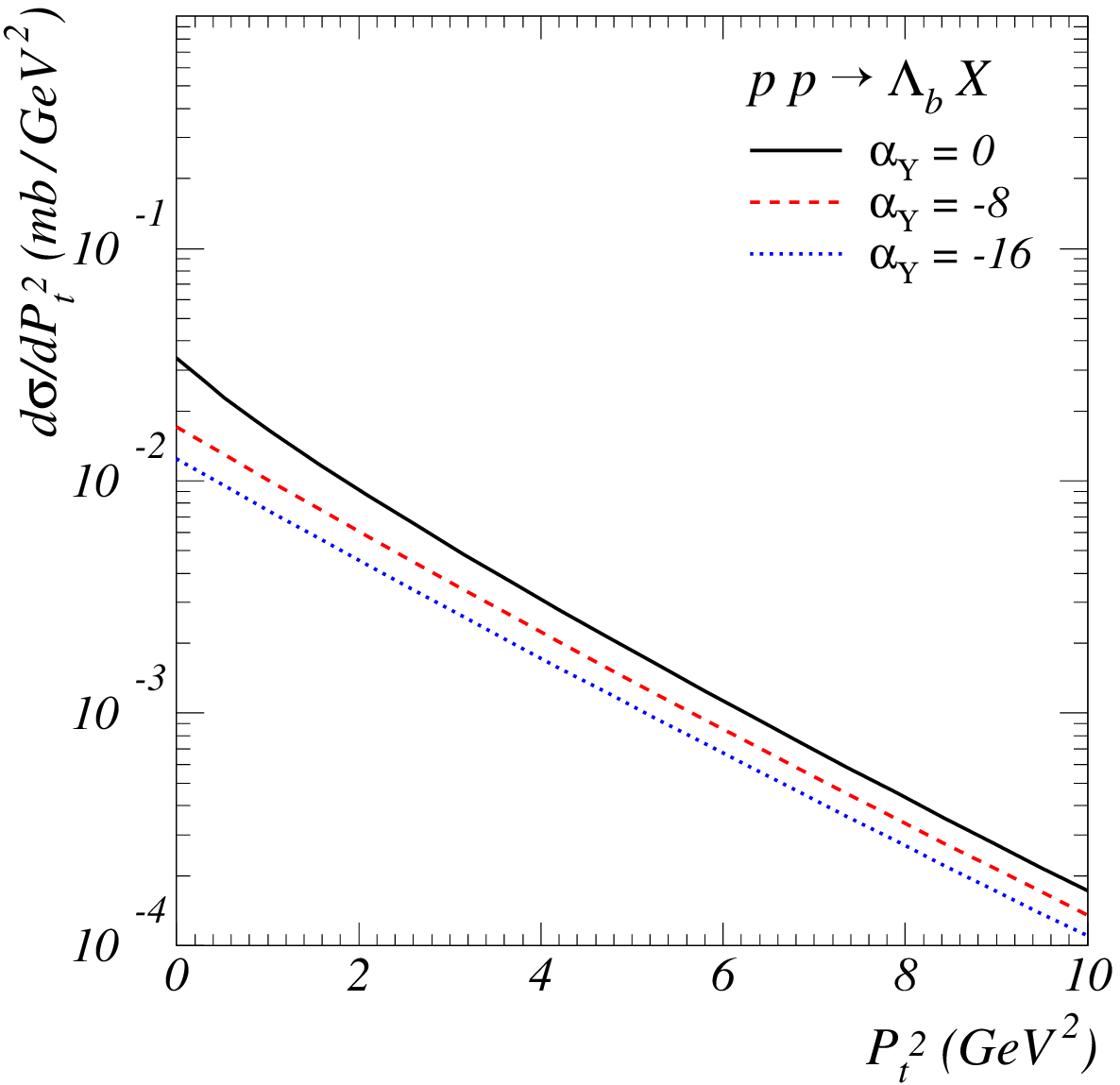,width=0.45\linewidth}}
\end{tabular}
\end{center}
 \caption[Fig.3]{The differential cross section $d\sigma/dx$ (left) and
$d\sigma/dP_t^2$ (right) for the
 inclusive process $pp\rightarrow\Lambda_b X$ at $\sqrt{s}=4~\mathrm{TeV}$.
}
\end{figure}
The $p_t$-inclusive spectrum of $\Lambda_b$ has much lower sensitivity to this quantity,
according to the results presented in Fig.3 (right).
Actually, our results presented in Fig.3 could be considered as some predictions for 
future experiments at LHC, see Fig.4.

Now let us analyze the production of the beauty hyperon, namely $\Lambda^0_b$,
at small scattering angles $\theta_{\Lambda^0_b}$ in the $pp$ collision at LHC energies.
This study would be reliable for the future forward experiments at LHC.
The produced $\Lambda^0_b$ baryon can decay as $\Lambda^0_b\rightarrow J/\Psi \Lambda^0$,
and $J/\Psi$ decays into $\mu^+\mu^-$, its branching ratio ($Br=\Gamma_j/\Gamma $) is $5.93\pm 0.06$
percent, or into $e^+e^-$ ($Br=5.93\pm 0.06\%$), whereas $\Lambda^0$ can decay into 
$p\pi^-$ ($Br=\Gamma_j/\Gamma=63.9\pm 05\%$), or into $n\pi^0$ ($Br=35.8\pm 0.5\%$),
see Fig.4.   
\begin{figure}[ht]
  \begin{center}
    \epsfig{file=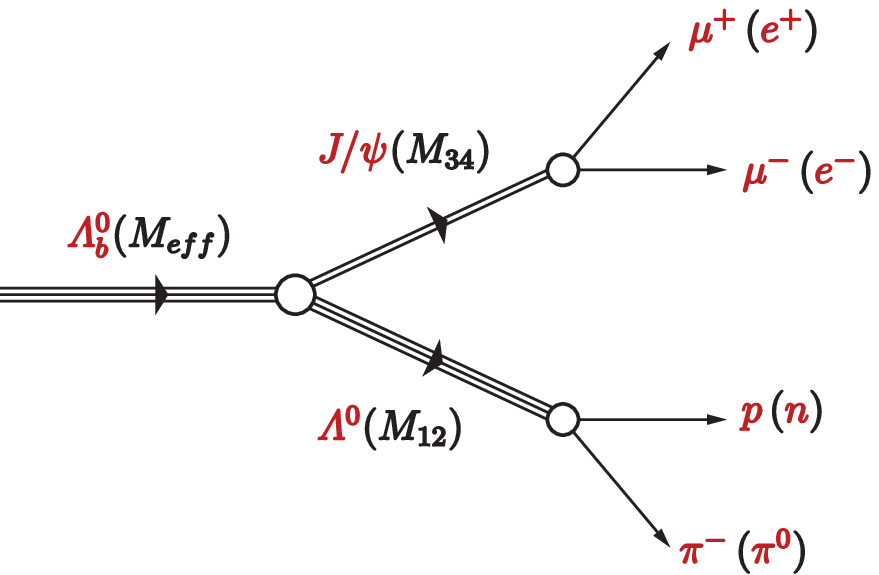,width=0.45\linewidth}
 \caption[Fig.4]{The decay $\Lambda_b\rightarrow J\Psi~\Lambda^0\rightarrow
\mu^+\mu^-(e^+e^-)~p\pi^-(n\pi^0)$.} 
 \end{center}
\end{figure}
\begin{eqnarray}
\frac{d\sigma}{d^3 p_1 dM_{34}}=\int\frac{d\sigma}{dM_{12} dM_{34}}\\
\nonumber
\delta^{(3)}({\bf p}_1+{\bf p}_2-{\bf p}_{12})dM_{12}~,
\end{eqnarray}
where 
\begin{eqnarray}
\frac{d\sigma}{dM_{12} dM_{34}}=
\nonumber
\int d^2p_{t\Lambda_b}
\frac{d\sigma_{pp\rightarrow\Lambda_b X}}{dx d^2p_{t\Lambda_b}} \\
\nonumber
Br_{\Lambda_b\rightarrow J/\Psi}
Br_{J/\Psi\rightarrow\mu^+\mu_-} Br_{\Lambda^0\rightarrow p\pi}
\frac{\pi^3}{2M^2_{eff}M_{12}M_{34}} \\
\nonumber
\lambda^{1/2}(M^2_{eff},M^2_{12}, M^2_{34})
\lambda^{1/2}(M^2_{12},M^2_1, M^2_2)\lambda^{1/2}(M^2_{34},M^2_3, M^2_4)~,
\end{eqnarray}
$Br_{\Lambda_b\rightarrow J/\Psi}=(4.7\pm 2.8)\cdot 10^{-4};~ 
Br_{J/\Psi\rightarrow\mu^+\mu_-}=(5.93\pm 0.06)\%;~
Br_{\Lambda^0\rightarrow p\pi}=(63.9\pm 0.5)\% $.\\
Here $\lambda(x^2,y^2,z^2)=((x^2-(y+z)^2)((x^2-(y-z)^2)$\\
One can get the following relation
\begin{equation}
d^3p_1~=~\frac{1}{2}p\xi_p d\phi_1 d\xi_p dt_p~,
\end{equation}
where $\xi_p=\Delta p/p$ is the energy loss, $t_p=(p_{in}-p_1)^2$ is the four-momentum transfer, 
$\phi_1$ is the azimuthal angle of
the final proton with the three-momentum ${\bf p}_1$.\\
Experimentally one can measure the differential cross section
\begin{equation}
\frac{d\sigma}{d\xi_p dt_p dM_{J/\Psi}}~=~\frac{1}{2}p\xi_p\int\frac{d\sigma}{d^3 p_1 dM_{34}}d\phi_1 
\end{equation}
This distribution could be reliable for the TOTEM experiment, where $J/\Psi$ decays into $\mu^+\mu^-$
and $\Lambda^0_b$ decays into $\pi^- p$ or for the ATLAS forward experiment, where $\Lambda^0_b$
decays as $\Lambda^0_b\rightarrow J/\Psi~\Lambda^0\rightarrow e^+e^-~\pi^0 n$ (Fig.4).
\begin{figure}[htb]
\begin{center}
\begin{tabular}{cc}
\mbox{\epsfig{file=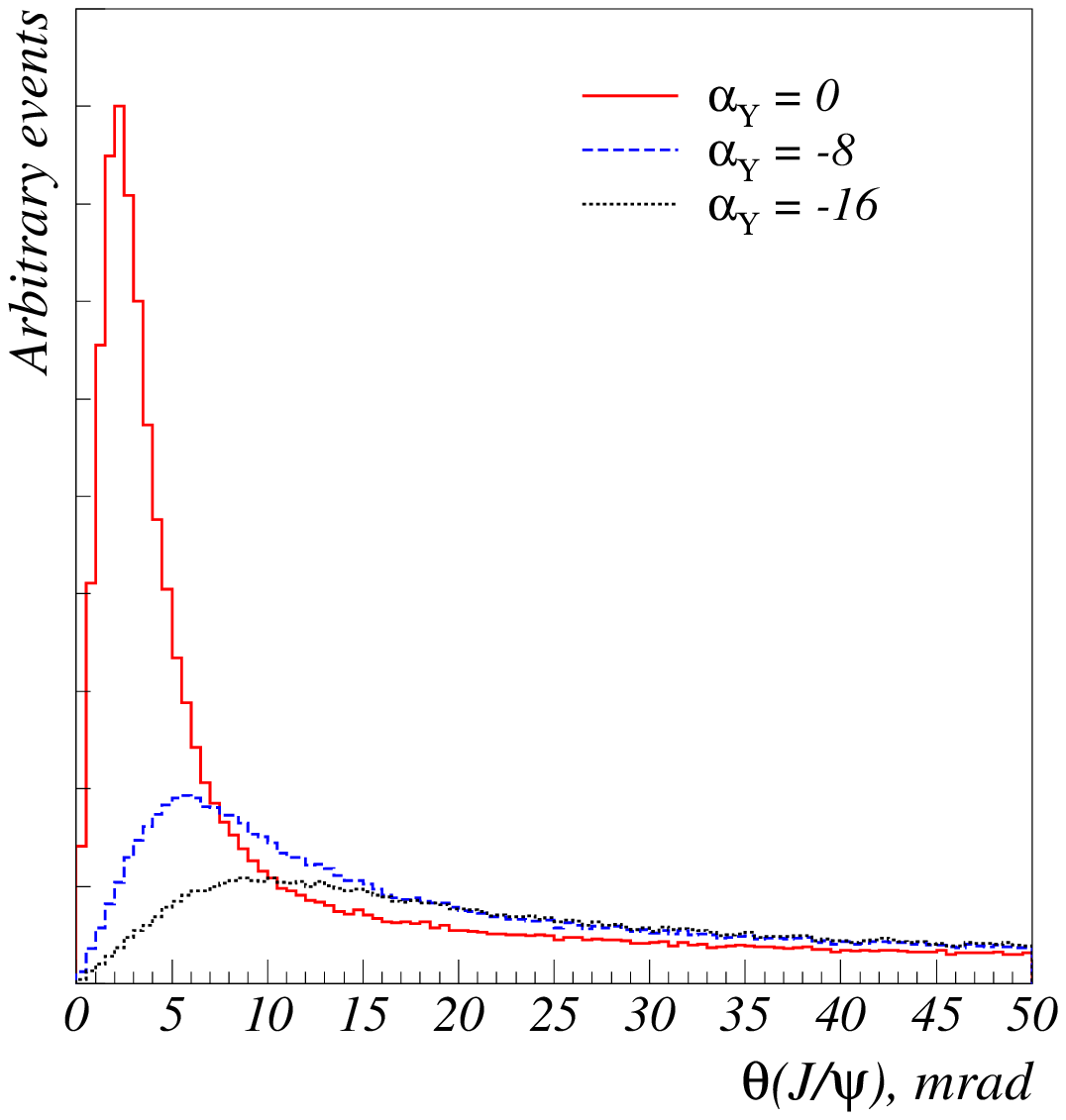,width=0.45\linewidth}} &
\mbox{\epsfig{file=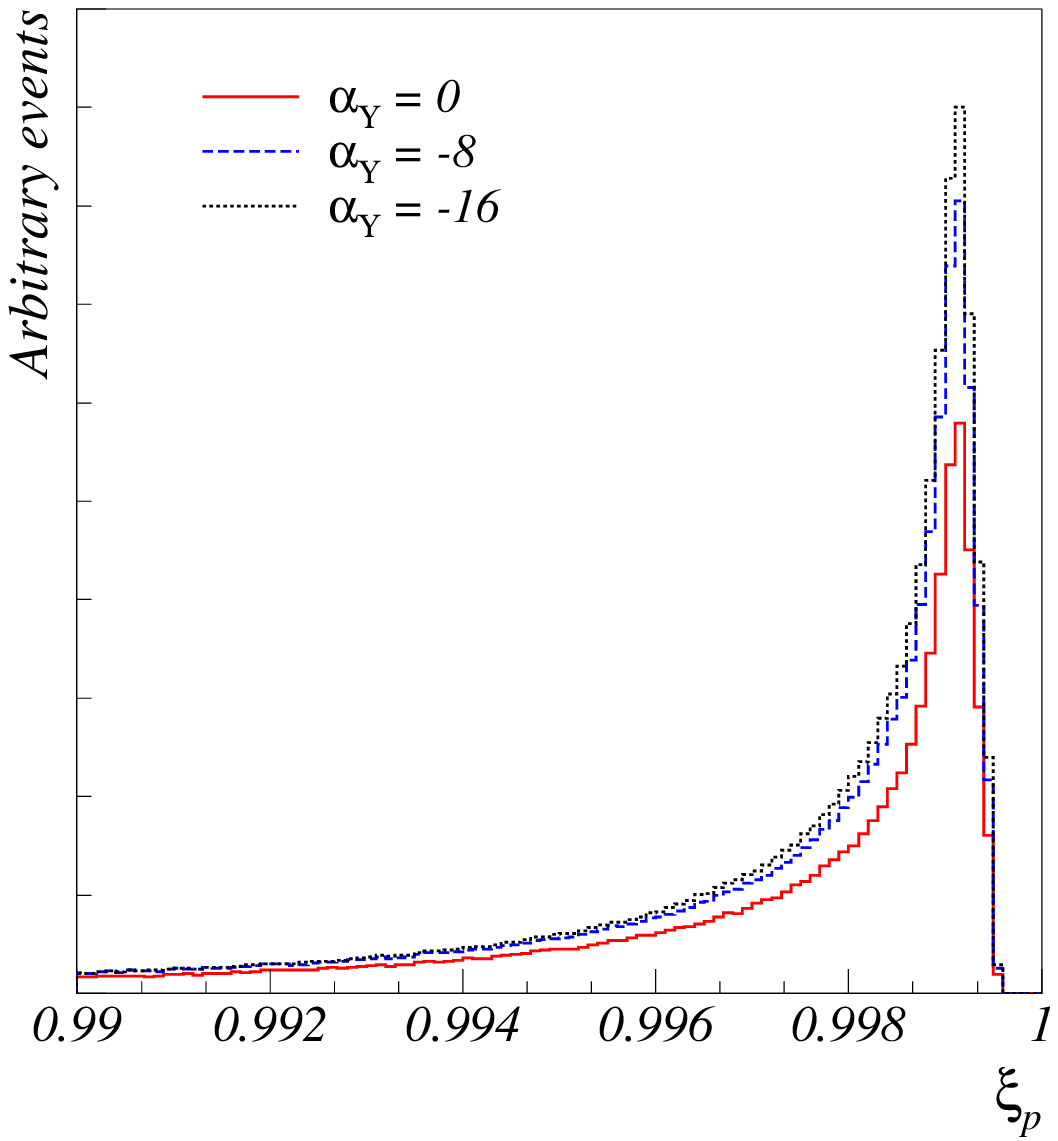,width=0.45\linewidth}}
\end{tabular}
\end{center}
 \caption[Fig.5]{The distributions over $\theta_{J/\Psi}$ (left) and $\xi_p$ (right) 
for the inclusive process 
$pp\rightarrow\Lambda_b X\rightarrow\mu^+\mu^- p\pi^- X$ at $\sqrt{s}=4.$ TeV
}
\end{figure}

In Fig.5 the  distributions over $\theta_{J/\Psi}$ (left) and $\xi_p$ (right) are presented 
at different values of the intercept $\alpha_\Upsilon(0)=0$ (solid line),
$\alpha_\Upsilon(0)=-8$ (dashed line) and $\alpha_\Upsilon(0)=-16$ (dotted line), 
where $\theta_{J/\Psi}$ is the scattering angle for the final $J/\Psi$. 
Fig.5 shows a sensitivity of these distributions to the intercept
of the $\alpha_\Upsilon$ Regge trajectory. Actually, the result presented in Fig.5 is 
a prediction for future LHC experiments on the heavy flavour baryon production at the LHC energies.    

\section{Conclusion}
 It was shown \cite{LKSB:09,LLB:09} that the modified QGSM including the
intrinsic longitudinal and transverse motion of quarks
(antiquarks) and diquarks in colliding protons allowed us to
describe rather satisfactorily the existing experimental data on
inclusive spectra of heavy hadrons produced in $pp$ and $p{\bar p}$ collisions   
It allows us to make some predictions for future LHC forward experiments on the beauty 
baryon production in $pp$ collisions which can give us new information on
the beauty quark distribution in the proton and very interesting information on
the Regge trajectories of ($b{\bar b}$) mesons.

\section{Acknowledgments}
 We thank  M. Deile,  K. Eggert, D. Elia, P. Garfstr\"{o}m, A. B. Kaidalov,  
A. D. Martin, M. Poghosyan and N. I. Zimin for very useful discussions. 
This work was supported in part by the RFBR grant N 08-02-01003.

\end{document}